# Valley-dependent topological interface states in biased armchair nanoribbons of gapless single-layer graphene for transport applications

**Zheng-Han Huang [1,*], Jing-Yuan Lai [2] and Yu-Shu Wu [1,2,3,*]**

[1] Institute of Electronic Engineering, National Tsing-Hua University, Hsin-Chu 30013, Taiwan
[2] College of Semiconductor Research, National Tsing-Hua University, Hsin-Chu 30013, Taiwan; laijeffery0301@gmail.com
[3] Department of Physics, National Tsing-Hua University, Hsin-Chu 30013, Taiwan
* Correspondence: fred200409@gapp.nthu.edu.tw (Z.-H.H.); yswu@ee.nthu.edu.tw (Y.-S.W.)

## Abstract

Valley-dependent topological physics offers a promising avenue for designing nanoscale devices based on gapless single-layer graphene. To demonstrate this potential, we investigate an electrical bias-controlled topological discontinuity in valley polarization within a two-segment armchair nanoribbon of gapless single-layer graphene. This discontinuity is created at the interface by applying opposite in-plane, transverse electrical biases to the two segments. An efficient tight-binding theoretical formulation is developed to calculate electron states in the structure. In a reference configuration, we obtain energy eigenvalues and probability distributions that feature interface-confined electron eigenstates induced by the topological discontinuity. Moreover, to elucidate the implications of interface confinement for electron transport, a modified configuration is introduced to transform the eigenstates into transport-active, quasi-localized ones. We show that such states result in Fano "anti-resonances" in transmission spectra. The resilience of these quasi-localized states and their associated Fano fingerprints is examined with respect to fluctuations. Finally, a proof-of-concept band-stop electron energy filter is presented, highlighting the potential of this confinement mechanism and, more broadly, valley-dependent topological physics in designing nanoscale devices in gapless single-layer graphene.

## 1. Introduction

From a device perspective, gapless single-layer graphene [1–3] serves as a crucial platform for realizing nanoscale electronic devices due to its high carrier mobility [4,5], excellent thermal properties [6], and demonstrated wafer-scale growth [7–11]. Proposed applications include spin qubits [12], high-frequency and thermally stable integrated circuits [13], and highly conductive, scalable on-chip interconnects utilizing graphene nanoribbons [14].

However, given their low-dimensional nature, the performance of graphene-based devices—especially at the nanoscale—is sensitive to fluctuations associated with defects, non-planarity, structural deviations, and graphene–substrate interactions [15,16]. While material engineering approaches, such as the use of a hexagonal boron nitride substrate [17], can mitigate some adverse effects, topological resilience offers an alternative strategy to counter fluctuations in device design [18]. This perspective motivates the present study.

It is well established that resilient, localized interface states—known as kink states—emerge when a topological invariant, such as the Chern number, varies spatially across a domain wall. With the rise of 2D materials, investigations into this topological phenomenon have entered a new realm. In the context of valleytronics [19–22], valley-dependent topological features—specifically valley Chern numbers (+1/−1 corresponding to Dirac valleys K/K' [23] and valley magnetic moments up/down [24])—in inversion symmetry-broken, gapped graphene [12,25–29] and transition metal dichalcogenides [30–35] provide an alternative mechanism for realizing this phenomenon.

Conceptually, valley Chern number domain walls in gapped graphene—both single-layer and bilayer—are quite often taken to be constructed by inverting the band gap across adjacent domains. The existence of corresponding interface states along such domain walls has been explicitly demonstrated in 2D systems [23,36–38], closely paralleling the boundary modes found in 2D and 3D topological insulators [39–42]. However, such a gap inversion-based approach encounters significant technical challenges, as summarized below.

As illustrated in Figure 1, for single-layer graphene, the approach requires the implementation of a staggered sublattice potential, with distinct on-site energies, for example, $\Delta$ and $-\Delta$, for A and B sublattices, respectively [36]. Theoretically, such a potential breaks inversion symmetry and opens a bulk graphene band gap of $2\Delta$ at Dirac points, with conduction (valence) band edge states composed predominantly of A (B) site orbitals when $\Delta > 0$ [28]. A spatial sign flip of $\Delta$ across the domains therefore switches the states between conduction and valence band edges, resulting in a spatial A-B band gap inversion across the domains. Moreover, as valley Chern number signs depend on band edge states [36], the flip generates a valley Chern number domain wall. However, experimentally realizing a staggered sublattice potential presents a nontrivial challenge, as it generally requires control and precision at the unit-cell scale.

On the other hand, in the case of AB-stacked bilayer graphene, it is relatively straightforward to break inversion symmetry, open a band gap, and create a valley Chern number domain wall. For example, by applying an out-of-plane electric field to the layers, an interlayer potential asymmetry is produced, which induces an effective staggered sublattice potential [23,28,37]. It then follows that a spatial reversal of the electric field inverts the gap and generates a domain wall [23]. However, the wafer-scale growth of AB-stacked bilayer graphene is currently somewhat challenging, as it is often hindered by the formation of domains with random twist angles [43,44].

In contrast to studies on gapped graphene, the present work focuses on gapless single-layer graphene ($\Delta = 0$), driven by its technological potential mentioned earlier. We consider simple, regular-edged armchair nanoribbons in gapless single-layer graphene subjected to in-plane, transverse electrical biases. The transverse size quantization in these graphene nanoribbons (GNRs) opens up energy gaps [45], rendering the structures semiconducting and thus suitable for the fabrication of nanoscale semiconductor devices [46–48]. Moreover, under in-plane, transverse electrical biases, these GNRs become inversion symmetry-broken, and their electron states exhibit bias polarity-dependent valley polarizations and magnetic moments [49,50]. Therefore, while such quasi-one-dimensional (Q1D) structures are not rigorously characterized by the valley Chern numbers of 2D layers, some manifestation of valley-dependent topological phenomena can still be anticipated. Exploring this topological physics may offer a viable pathway toward the practical implementation of devices based on such structures.

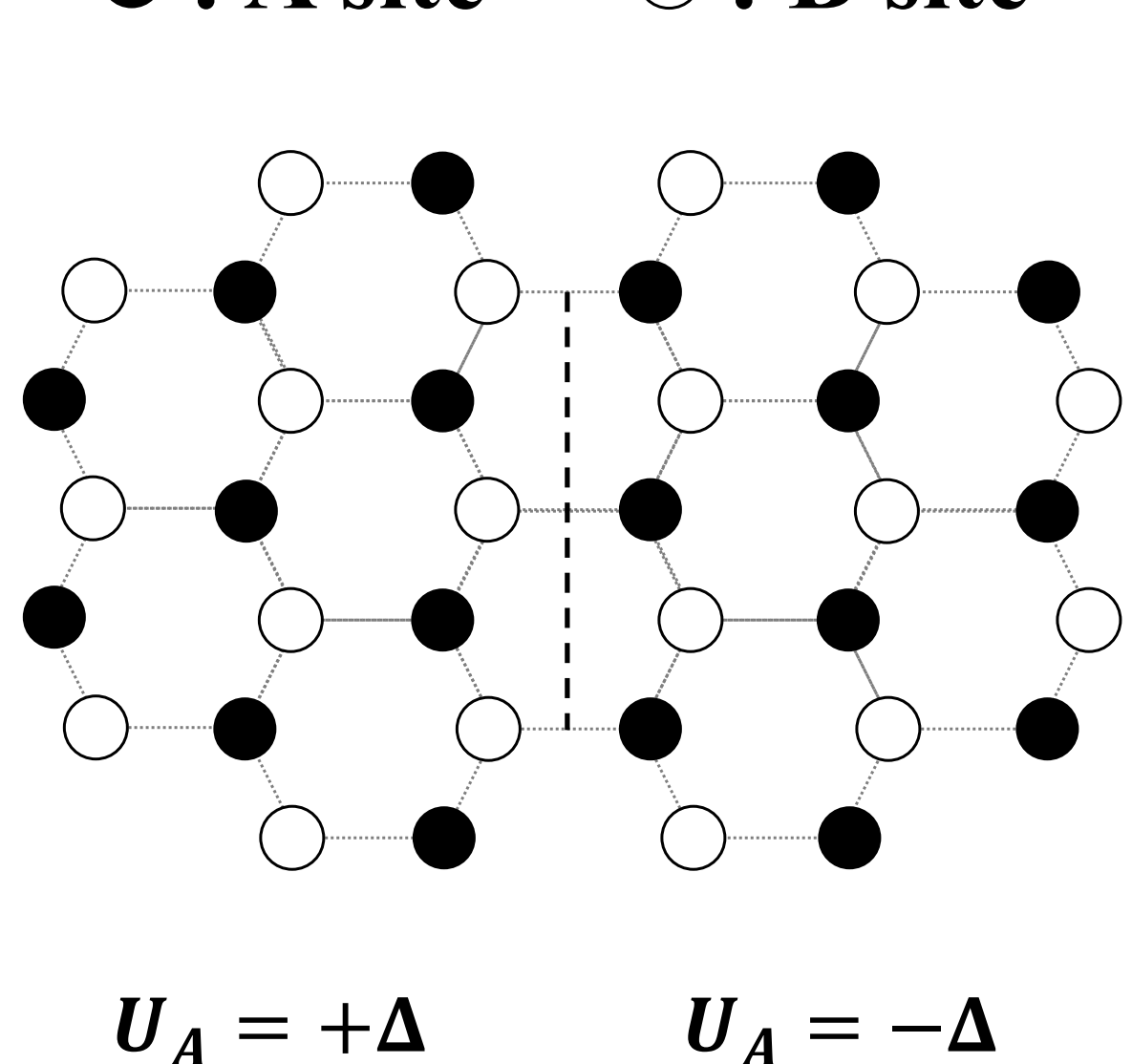

**Figure 1.** A valley Chern number domain wall in gapped single-layer graphene. The A (black) and B (white) sublattice sites have on-site energies $U_A = +\Delta$ and $U_B = -\Delta$ in the left domain, and $U_A = -\Delta$ and $U_B = +\Delta$ in the right domain. This produces a valley Chern number domain wall, indicated by the black dashed vertical line.

Specifically, we investigate an electrical bias-controlled discontinuity in valley polarization within a two-segment armchair nanoribbon. The discontinuity is created by applying opposite in-plane, transverse electrical biases to the two segments. Unlike the gap inversion-based approach [23,36,37], we demonstrate a mechanism that can induce a discontinuity of topological character without relying on a band gap or staggered sublattice potential.

Figure 2 provides an explanation of the mechanism. It presents a two-segment, source/drain structure, formed of an armchair GNR and designated as Structure X throughout this work, where the source and drain electrodes are subjected to transverse, linear bias potentials of opposite signs. The armchair edge runs along the longitudinal ($x$) direction, and the Dirac wave vectors **K** and **K'** run along the transverse (y) direction. Generally, as scattering off armchair edges takes **K** (**K'**) to **K'** (**K**), the two Dirac valleys are coupled, and an electron state in the nanoribbon exhibits both valley components.

The weights of the two components in the state depend on the bias, as explained below. Let $k_x$ be the electron wave vector. In the absence of bias, the structure is mirror-symmetric with respect to the operation y → −y, so the scattering mixes the valleys with equal weights [45], resulting in a vanishing valley polarization and magnetic moment. The application of a transverse, in-plane electrical bias breaks the mirror symmetry, introduces an asymmetry between the two valleys, and thus valley-polarizes the electron, with the resultant polarization and orbital magnetic moment exhibiting a dependence on both $k_x$ and the bias polarity [49,50]. Therefore, applying biases of opposite polarity to the two electrodes induces a valley polarization discontinuity of topological origin at the interface, as depicted in Figure 2. It is of fundamental interest to investigate this discontinuity and explore whether the present system shares essential features with the aforementioned well-studied topological systems characterized by valley Chern numbers—specifically, electron confinement near the discontinuity.

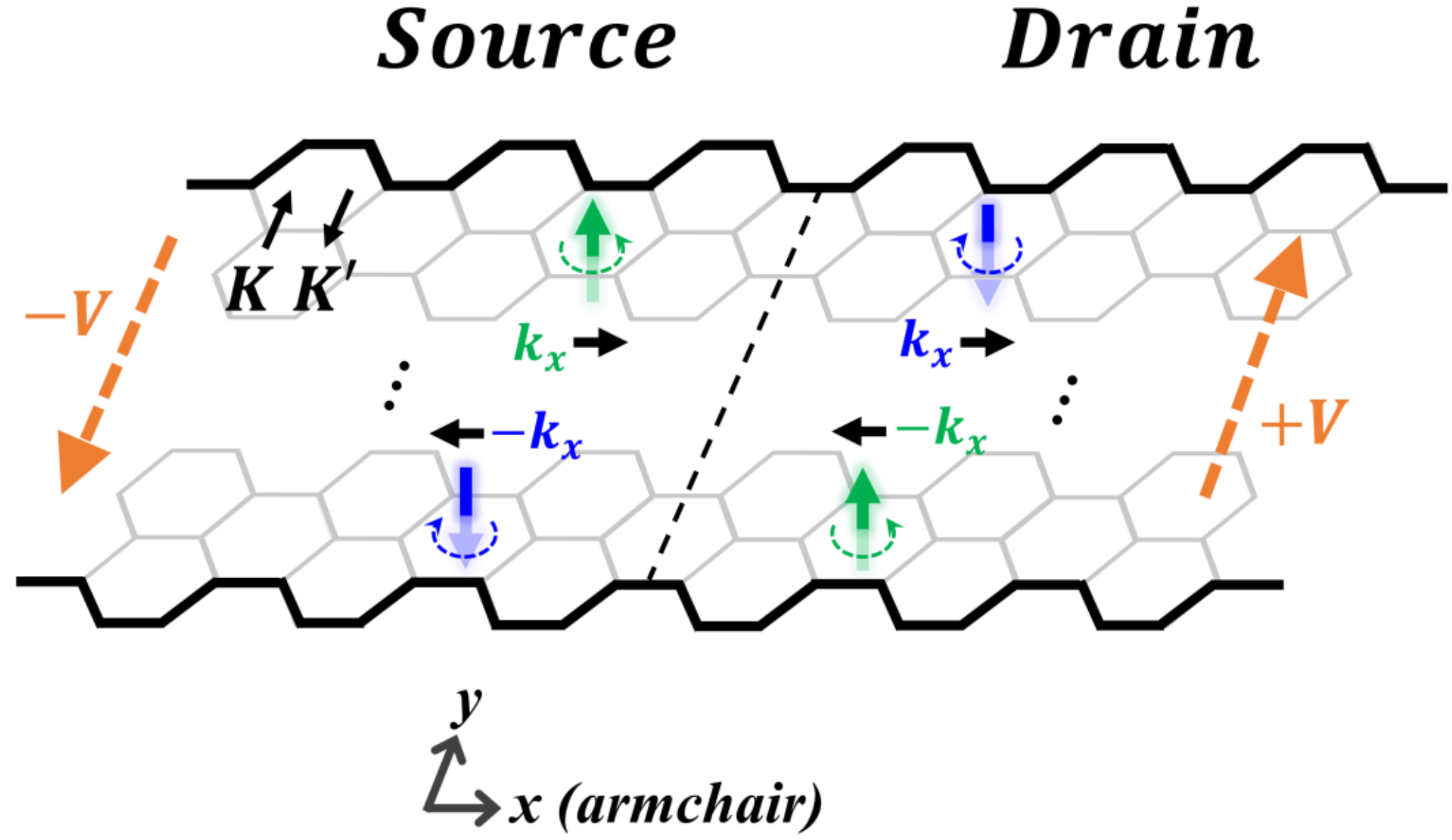

**Figure 2.** A valley-dependent domain wall in Structure X, which is a source/drain structure of armchair nanoribbon in gapless single-layer graphene, where the electrodes are subjected to biases −V and +V (orange arrows), respectively. For the electron at a given wave vector ($k_x$, or $-k_x$), these biases generate opposite valley polarizations and opposite valley magnetic moments (green and blue arrows) in the electrodes, inducing a discontinuity of topological kind at the interface (black dashed line). The black arrows labeled **K** and **K'** in the upper left illustrate the valley coupling mechanism, as scattering off the armchair edges mixes the **K** and **K'** valley components of the electron states.

In passing, we note that in non-valleytronic contexts, the phenomenon of kink states has also been explored, particularly in Q1D structures of gapless single-layer graphene [51]. For example, Chou et al.

investigated inversion-symmetric nanoribbon structures with nontrivial edge termination patterns [52]. In particular, they calculated topological $Z_2$ invariants and demonstrated the existence of end states at the terminations of certain nanoribbon segments, as well as junction states between segments possessing distinct $Z_2$ invariants.

In our investigation, we begin by examining and confirming the interface confinement behavior in Structure X. Next, we explore the implications of this confinement for electron transport, with a focus on potential transport device applications. Finally, a proof-of-concept band-stop electron energy filter is presented to highlight the potential of such confinement and, more broadly, valley-dependent topological physics in designing nanoscale devices in gapless single-layer graphene.

The paper is organized as follows. In Section 2, a theoretical formulation is presented for calculating interface eigenstates. An efficient method involving only a single unit cell is developed. In addition, a recursive Green's function algorithm is sketched for the calculation of electron transmission. Section 3 presents the results of interface eigenstates in Structure X. Section 4 investigates the impact of interface confinement on electron transport. A configuration modified from Structure X is introduced to transform the interface-localized eigenstates into transport-active, quasi-localized ones. The effects of these quasi-localized states on transmission are illustrated, and their resilience is examined with respect to configurational fluctuations. A band-stop electron energy filter is discussed to illustrate the possible utilization of quasi-localized interface states for transport applications. Finally, Section 5 summarizes the main findings.

## 2. Theoretical Methods

To investigate the topological interface depicted in Figure 2, we establish a formulation to calculate interface states in a discrete, gapless single-layer graphene lattice-based structure. Specifically, for this work, we focus on a geometry consisting of two identical, semi-infinite nanoribbon segments connected back-to-back, aligned along the *x*-axis and subject to transverse, in-plane bias-induced potentials with opposite signs for the two segments.

Generally, the tight-binding method provides a reasonable theoretical description of energy band structures [53]. In addition, in the field of quantum transport, the well-established framework of the recursive Green's function (RGF) method developed by T. Ando [54] routinely employs the tight-binding model. Therefore, in our study, we apply the tight-binding model in both calculations of energy band structure and RGF.

Within the tight-binding framework, the Hamiltonian is explicitly defined as follows:

$$H = -t\sum_{<m,m'>} c_m^{\dagger} c_{m'} + \sum_m U(x_m, y_m) c_m^{\dagger} c_m, \tag{1}$$

where $c_m$ and $c_m^{\dagger}$ are, respectively, the lowering and raising operators of the atomic $2p_z$ orbital on site $m$, $<m, m'>$ denotes a pair of nearest neighbor sites, $t$ is the corresponding nearest neighbor hopping parameter, and the second term in $H$ describes the bias-induced on-site energy shift, e.g., $U(x_m, y_m) = \text{sgn}(x_m)\, V(y_m)$, with $V(y_m)$ linear in $y_m$, where $\text{sgn}(x_m)$ is the sign of $x_m$ and $(x_m, y_m)$ denotes the position of site $m$.

We extend each segment to an infinite nanoribbon and compute its complex band structure, $E(k_x)$, and nanoribbon states, where E is the electron energy and $k_x$ is the Bloch wave vector. As is standard in band structure calculations, considering a single nanoribbon unit cell suffices. A method for this computation was previously developed for 3D zinc-sulfide crystals [55,56] and is extended here to nanoribbons in 2D hexagonal crystals. In summary, for a given E, one computes both the wave vector $k_x^{(n)}$'s and the corresponding nanoribbon states $\psi_n$'s, where $n$ is the energy band index. Generally, $k_x^{(n)}$ is complex, with the imaginary part $\text{Im}(k_x^{(n)})$ vanishing for an extended state and non-vanishing for a state that decays exponentially. For each side of the interface (at $x = 0$), the nanoribbon state obtained for E and $k_x^{(n)}$ is expressed as follows:

$$\psi_n^{(l)} = \Sigma_m \psi_{n,m}^{(l)} \phi_m, \tag{2}$$

where $l \in \{S, D\}$ labels the two sides of the interface, designated as source (S) for $x < 0$ and drain (D) for $x > 0$ throughout the work; $\phi_m$ denotes the $2p_z$ orbital on site $m$; and $\psi_{n,m}^{(l)}$ is the projected amplitude of $\psi_n^{(l)}$ on site $m$.

Let $\Psi$ be the interface state at energy E. We construct it as a linear combination of nanoribbon states for each side of the interface. To ensure the spatial localization of $\Psi$ near the interface, only those nanoribbon states that exhibit exponential decay away from the interface are included in the expansion. Let N denote the total number of such states for either side of the interface. Explicitly, the corresponding amplitude of $\Psi$ on site $m$ is given by the following:

$$\Psi_m^{(l)} = \Sigma_{n=1}^{N} c_n^{(l)} \psi_{n,m}^{(l)} \qquad (l = S \text{ or } D), \tag{3}$$

where $c_n^{(l)}$'s are the coefficients of linear combinations.

We enforce the continuity of $\Psi$ across the interface. As a representative example, we consider the interface section in a simple source/drain structure illustrated in Figure 3, which is, from left to right, composed of hexagonal blocks in the 2–1–2 sequence, with the interface cutting through the hexagon in the middle. The continuity equations are given by $\Psi_{B3}^{(S)} = \Psi_{B3}^{(D)}$, $\Psi_{A2'}^{(S)} = \Psi_{A2'}^{(D)}$, and similarly for sites B7 and A6′. We note that these equations imply the continuity of both bulk cell probability, $\rho_{cell}^{(S)} = \rho_{cell}^{(D)}$, and bulk cell current continuity, $j_{cell}^{(S)} = j_{cell}^{(D)}$. Specifically, for the bulk unit cell comprising sites B3 and A2′, these quantities are given by $\rho_{cell}^{(S)}(\mathrm{B3} \leftrightarrow \mathrm{A2'}) = \left|\Psi_{B3}^{(S)}\right|^2 + \left|\Psi_{A2'}^{(S)}\right|^2$ and $j_{cell}^{(S)}(\mathrm{B3} \leftrightarrow \mathrm{A2'}) = -it\left({\Psi_{B3}^{(S)}}^{\dagger}\Psi_{A2'}^{(S)} - {\Psi_{A2'}^{(S)}}^{\dagger}\Psi_{B3}^{(S)}\right)$. Analogous relations hold for the other bulk cell consisting of B7 and A6′.

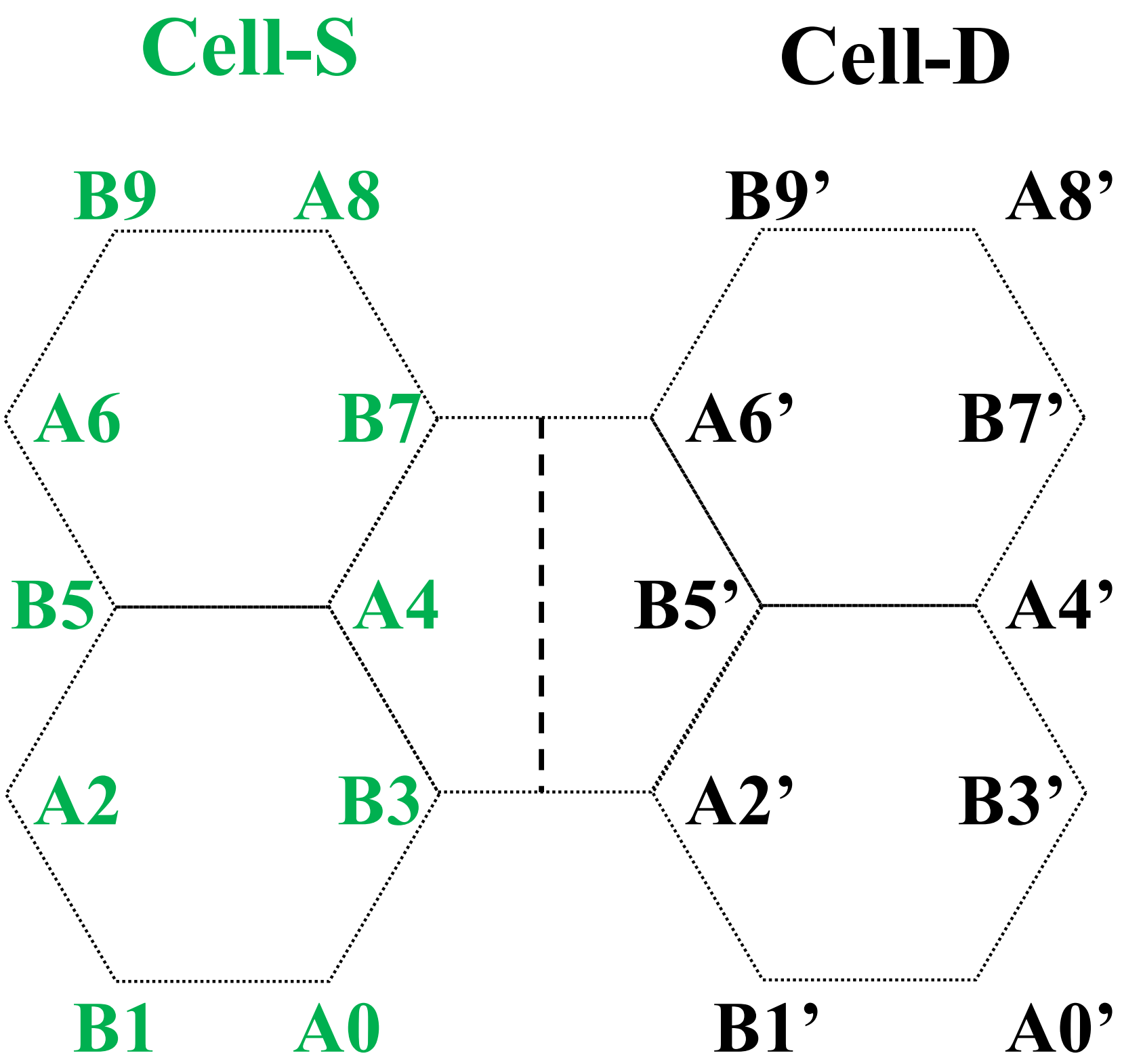

**Figure 3.** Schematic of a cross-section near the source/drain interface (black dashed vertical line) for a narrow armchair graphene nanoribbon with a width of 2 hexagons per nanoribbon unit cell. Cell-S and Cell-D denote the nanoribbon unit cells on the source and drain sides, respectively. Each unit cell consists of alternating A and B sublattice sites, as labeled.

Wave function continuity equations for the structure in Figure 3 can be organized into the following matrix equation:

$$M \cdot C = 0,$$

$$: \begin{pmatrix} \psi^{(S)}_{1,A2'} & \psi^{(S)}_{2,A2'} & \psi^{(D)}_{1,A2'} & \psi^{(D)}_{2,A2'} \\ \psi^{(S)}_{1,B3} & \psi^{(S)}_{2,B3} & \psi^{(D)}_{1,B3} & \psi^{(D)}_{2,B3} \\ \psi^{(S)}_{1,A6'} & \psi^{(S)}_{2,A6'} & \psi^{(D)}_{1,A6'} & \psi^{(D)}_{2,A6'} \\ \psi^{(S)}_{1,B7} & \psi^{(S)}_{2,B7} & \psi^{(D)}_{1,B7} & \psi^{(D)}_{2,B7} \end{pmatrix}, \ C = \begin{pmatrix} c_1^{(S)} \\ c_2^{(S)} \\ -c_1^{(D)} \\ -c_2^{(D)} \end{pmatrix}. \tag{4}$$

Solving the foregoing equation yields the solution of coefficients $c_1^{(S)}, c_2^{(S)}, c_1^{(D)}$, and $c_2^{(D)}$. When the determinant of M vanishes, i.e.,

$$\det(M) = 0, \tag{5}$$

the solution obtained becomes nontrivial, indicating the presence of an interface state at energy E. The above formulation can be readily generalized to nanoribbons of arbitrary width, including those in the limit of divergent width, i.e., 2D structures consisting of semi-infinite planes of source and drain.

While formulated for a discrete lattice, one can show that, in the continuum limit, this reduces to the continuity of $(\Psi_A(x,y), \Psi_B(x,y))$ at $x = 0$, where $(\Psi_A, \Psi_B)$ is the Dirac two-component wave function defined in the continuum *xy*-plane. We note that this agrees with the previous work in the continuum limit, for example, that of Morpurgo and co-workers, who studied chiral zero modes in a 2D bilayer graphene structure and also derived wave-function continuity [37].

Finally, we outline the procedure for calculating electron transmission. We denote the wave functions on the source and drain sides as $\Psi^{(S)}$ and $\Psi^{(D)}$, respectively, given by the following:

$$\Psi^{(S)} = \Psi^{(S)}_{inc} + \Psi^{(S)}_{ref}, and\ \Psi^{(D)} = \Psi^{(D)}_{trans}, \tag{6}$$

where $\Psi^{(S)}_{\text{inc}}$, $\Psi^{(S)}_{\text{ref}}$, and $\Psi^{(D)}_{\text{trans}}$ represent the incident, reflected, and transmitted state, respectively. The incident state $\Psi^{(S)}_{inc}$ corresponds to an extended, forward-moving nanoribbon state $\psi^{(S)}_{n=i}$ with $k_x^{(i)} > 0$. The reflected state $\Psi^{(S)}_{\text{ref}}$ is a linear combination of both extended, backward-moving nanoribbon states ($\psi^{(S)}_{n=n'}$ with $k_x^{(n')} < 0$) and those exponentially decreasing away from the interface ($\psi^{(S)}_{n=n''}$ with $\text{Im}(k_x^{(n'')}) < 0$). Explicitly, the projected amplitude on site *m* is $\Psi^{(S)}_{\text{ref},m} = \Sigma_{n\in\{n's,n''s\}} c_n^{(S)} \psi^{(S)}_{n,m}$, where the $c_n^{(S)}$'s are the expansion coefficients. Similarly, the transmitted state $\Psi^{(D)}_{\text{trans}}$ is composed of linear combinations of extended, forward-moving nanoribbon states ($\psi^{(D)}_{n=n'}$ with $k_x^{(n')} > 0$) and exponentially decaying states ($\psi^{(D)}_{n=n''}$ with $\text{Im}(k_x^{(n'')}) > 0$). Explicitly, its projected amplitude on site *m* is $\Psi^{(D)}_{\text{trans},m} = \Sigma_{n\in\{n's,n''s\}} c_n^{(D)} \psi^{(D)}_{n,m}$, where the $c_n^{(D)}$'s are the expansion coefficients. Using the recursive Green's function (RGF) algorithm, we take the incident state site amplitudes $\psi^{(S)}_{i,m}$'s as input to calculate the coefficients $c_n^{(S)}$'s and $c_n^{(D)}$'s for the reflected and transmitted states in a recursive fashion [54]. The calculation yields the transmission coefficient T = $\Sigma_{n\in\{n's\}} \left|c_n^{(D)}\right|^2 J_n^{(D)}/J_i^{(S)}$. Here, $J_i^{(S)}$ and $J_n^{(D)}$ denote the total currents of nanoribbon states $\psi_i^{(S)}$ and $\psi_n^{(D)}$, respectively. For the structure in Figure 3, for example, $J_i^{(S)} = j^{(S)}_{i,cell}(\text{B3} \leftrightarrow \text{A2}') + j^{(S)}_{i,cell}(\text{B7} \leftrightarrow \text{A6}')$, where $j^{(S)}_{i,cell}(\text{B3} \leftrightarrow \text{A2}')$ and $j^{(S)}_{i,cell}(\text{B7} \leftrightarrow \text{A6}')$ are the bulk cell currents given earlier except for the replacement of $\Psi^{(S)}$ with $\psi_i^{(S)}$ in the current expression. Note that, in the expression for T, only extended components (with band index $n\in\{n's\}$) of the drain side are included, as $J_n^{(D)} = 0$ for exponentially decaying components (with band index $n\in\{n''s\}$).

## 3. Interface States

We designate Structure X as shown in Figure 2 as the reference configuration and apply the theoretical framework of Section 2 to examine its interface states. In this study, the structure is taken to be 18 hexagons wide in the y-direction. The bias-induced linear potential, V(y), has an overall amplitude of 0.35 eV across the transverse dimension.

The results are presented in Figure 4. Figure 4a shows the nanoribbon energy band structure for both the source and drain, covering three conduction bands and one valence band. The first conduction band edge (CBE-1) is located at 0.135 eV and the first valence band edge (VBE-1) at −0.135 eV, demonstrating the electron-hole symmetry. It is worth noting that the source and drain band structures are identical based on the two following facts: (1) the transformation V(y) ⟶ V(−y) maps the source potential to the drain potential and vice versa, giving E($k_x$, source) = E(−$k_x$, drain); and (2) due to the mirror reflection symmetry under $x \longrightarrow -x$, E($k_x$) = E(−$k_x$) in both the source and drain.

Figure 4b displays the corresponding complex band structure used in the construction of matrix M in Equation (4). Black curves represent energy bands for real $k_x$, while red and green curves together illustrate complex energy bands where Im($k_x$) ≠ 0. Notably, for a state with both real and imaginary parts of $k_x$ non-vanishing, there exists a four-fold degeneracy with wave vectors ($k_x$)*, −$k_x$, and (−$k_x$)* due to time reversal symmetry and the mirror reflection symmetry under $x \longrightarrow -x$. In this case, a curve of E vs. |Re($k_x$)| or E vs. −|Im($k_x$)| represents four degenerate states.

Figure 4c plots log|det(M)|. As the condition det(M) = 0 determines interface states, the dips at $E_0$, $E_1$, …, in the plot are identified as interface state energy eigenvalues. Several discontinuities appear in the plot. Each derives from a corresponding abrupt change in the dimension of M that occurs when E crosses a band edge, since the number of exponentially decaying states used to construct M varies discontinuously, as indicated in Figure 4b.

We now examine the two interface state solutions at $E_0$ = 0.098 eV and $E_1$ = 0.23 eV. The state at $E_0$ lies within the nanoribbon band gap, while that at $E_1$ resides within the first conduction band. The 1D probability distributions, $\rho_0(x)$ and $\rho_1(x)$, of the two states, respectively, are presented in Figure 4d. Each distribution clearly reveals a pronounced peak at the interface ($x$ = 0), confirming its interface-localized nature.

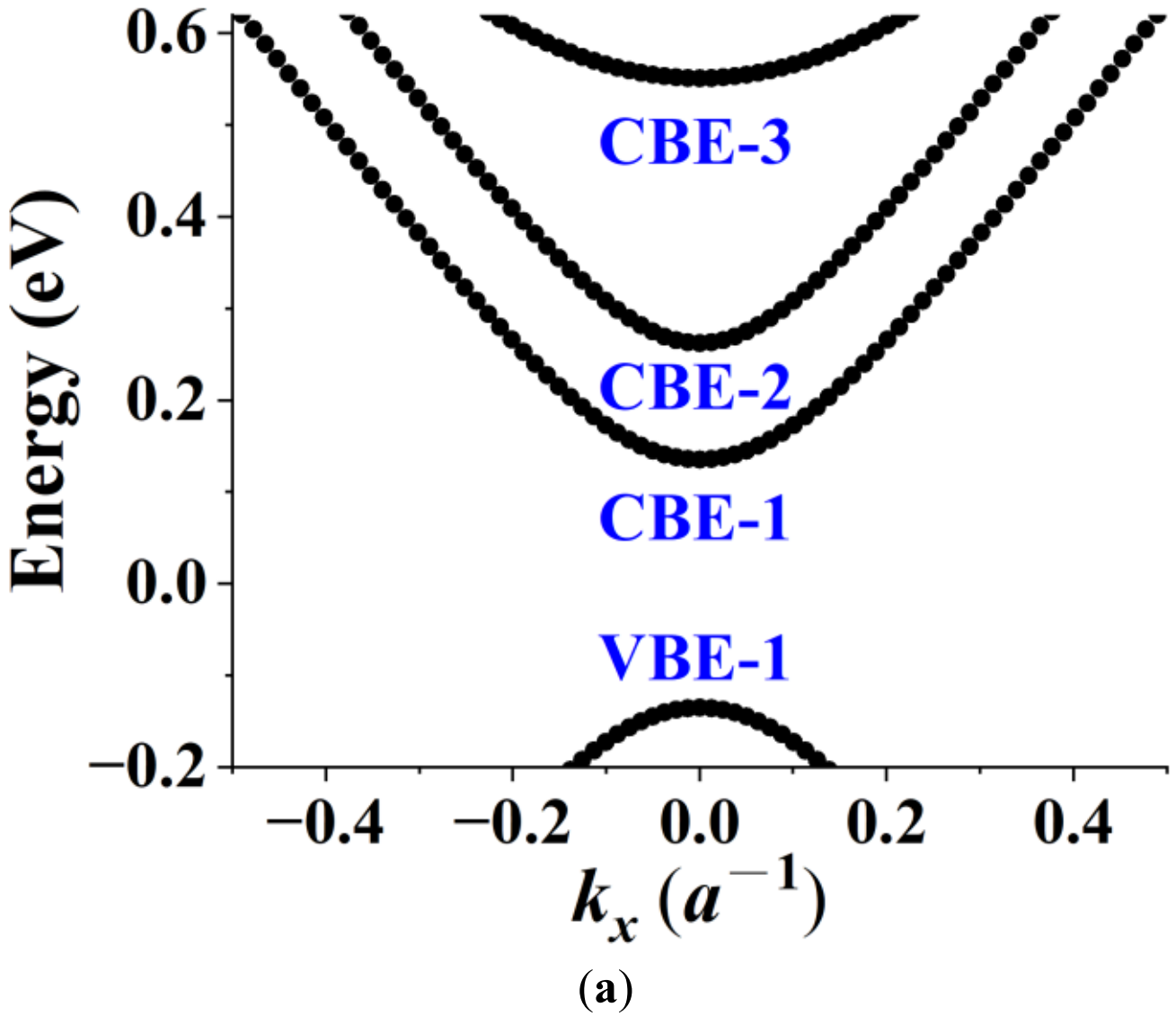

(a)

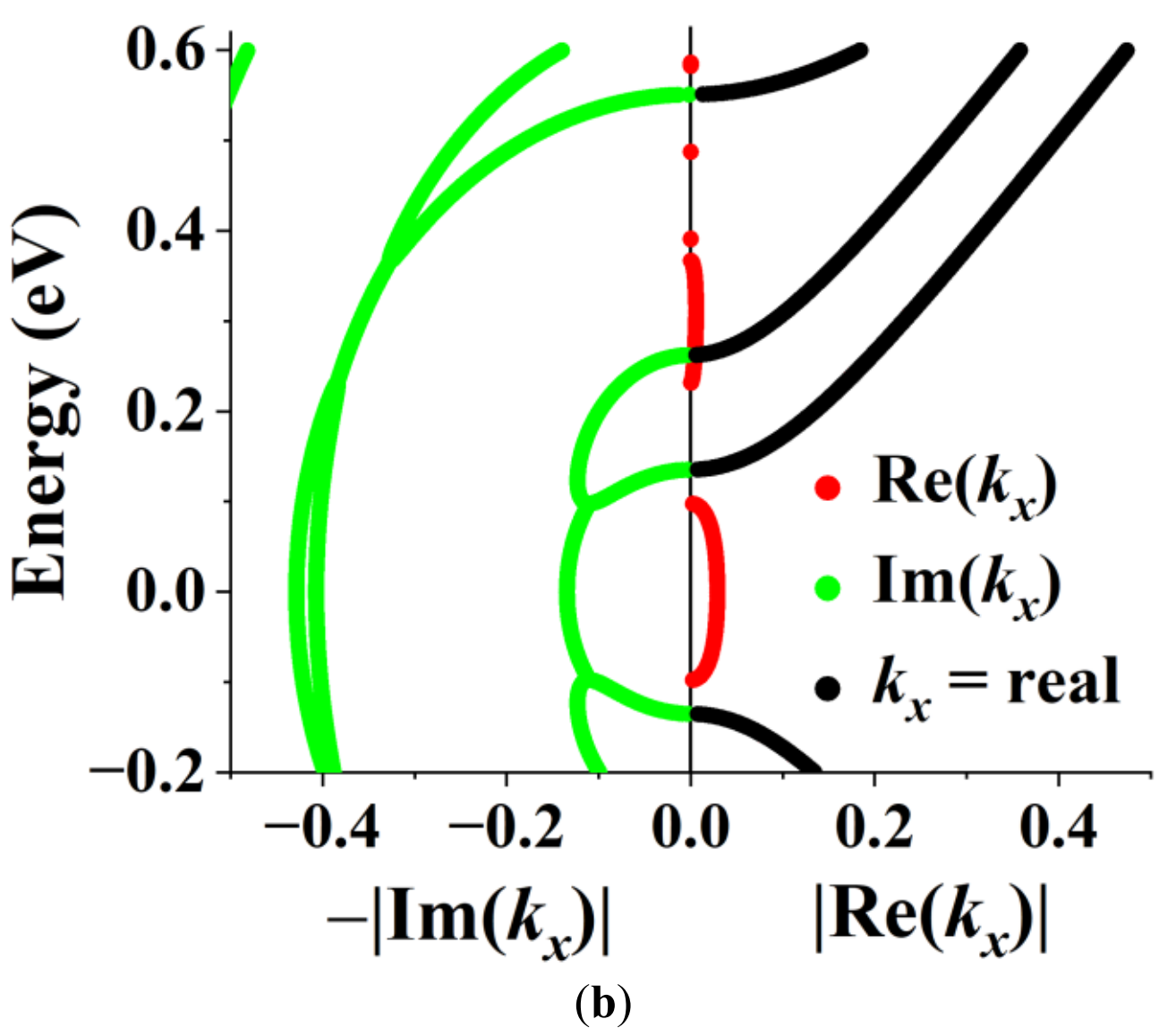

(b)

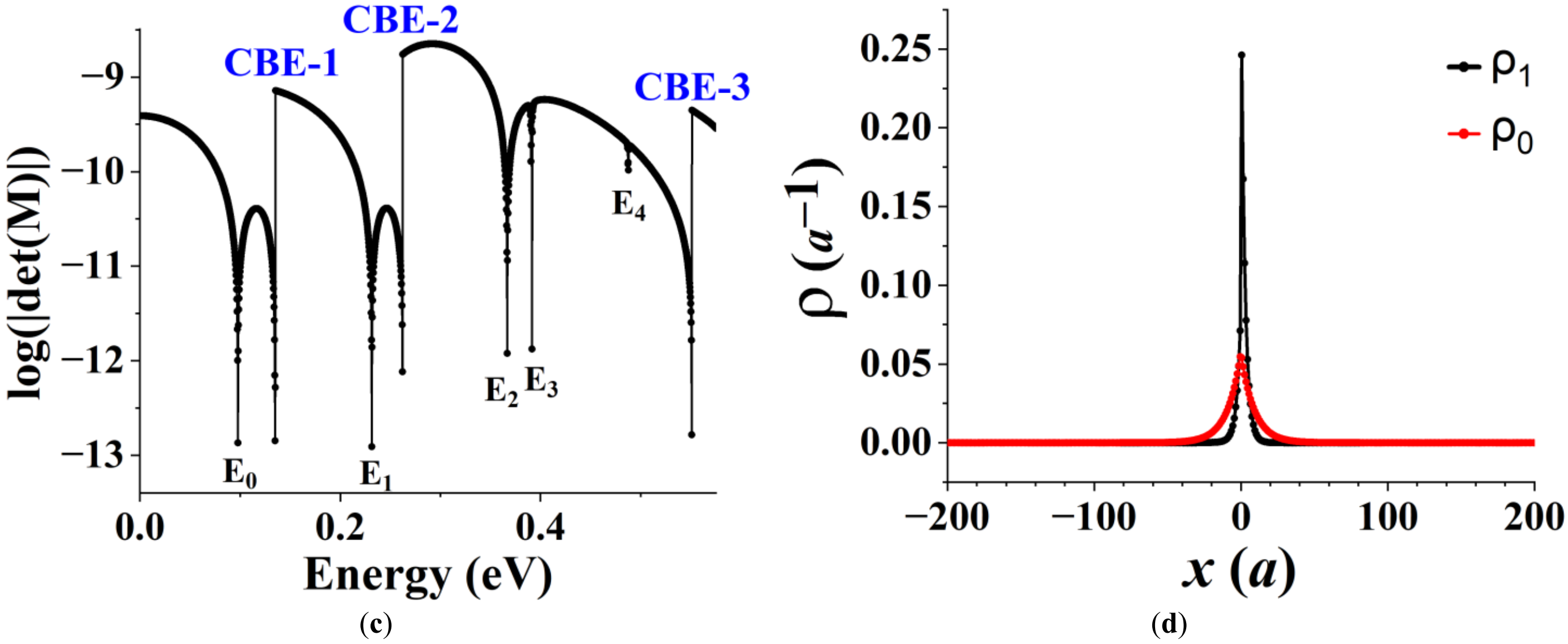

**Figure 4.** (**a**) Nanoribbon band structure of the source and drain in Structure X, showing three conduction bands and one valence band. The wave vector $k_x$ is expressed in units of $a^{-1}$, where $a = 3a_{cc}$ is the nanoribbon lattice constant and $a_{cc}$ is the carbon–carbon bond length. (**b**) Corresponding complex band structure. The absolute value of the real part of $k_x$, $|\mathrm{Re}(k_x)|$, is plotted on the right, and the negative absolute value of the imaginary part, $-|\mathrm{Im}(k_x)|$, on the left. Black curves represent real energy bands ($k_x$ is real), while red and green curves together illustrate complex energy bands ($\mathrm{Im}(k_x) \neq 0$). States at a given E are degenerate due to time-reversal and mirror-reflection symmetries. (**c**) Plot of log｜det(M)｜ vs. E. The dips at $E_0$, $E_1$, etc., identify the energy eigenvalues of interface states. Discontinuities marked CBE-1, CBE-2, etc., indicate band edges, reflecting abrupt changes in the number of exponentially decaying states used to construct M. (**d**) One-dimensional probability distributions $\rho_0(x)$ and $\rho_1(x)$, each defined as electron probability per nanoribbon unit cell, for the states at $E_0$ = 0.098 eV and $E_1$ = 0.23 eV, respectively. Both distributions show pronounced localization at the interface ($x$ = 0).

## 4. Effects on Electron Transport

In Structure X, since interface states are localized eigenstates, they are completely decoupled from the extended scattering states, specifically the incident, reflected, and transmitted states, used to describe electron transmission. This complete decoupling implies that electron transmission in Structure X cannot manifest any signature of interface states, as verified by the numerical result in Figure 5. In the figure, the transmission spectrum, T vs. E, calculated using incident states in the first conduction band, displays a step function-like curve. T sets on at CBE-1 and rapidly approaches unity, with occasional dips occurring when E crosses a band edge; these transitions alter the coupling between incident and transmitted states, thereby modulating the transmission T. As explicitly illustrated, at interface state energies $E_1$, $E_2$, etc., T remains featureless.

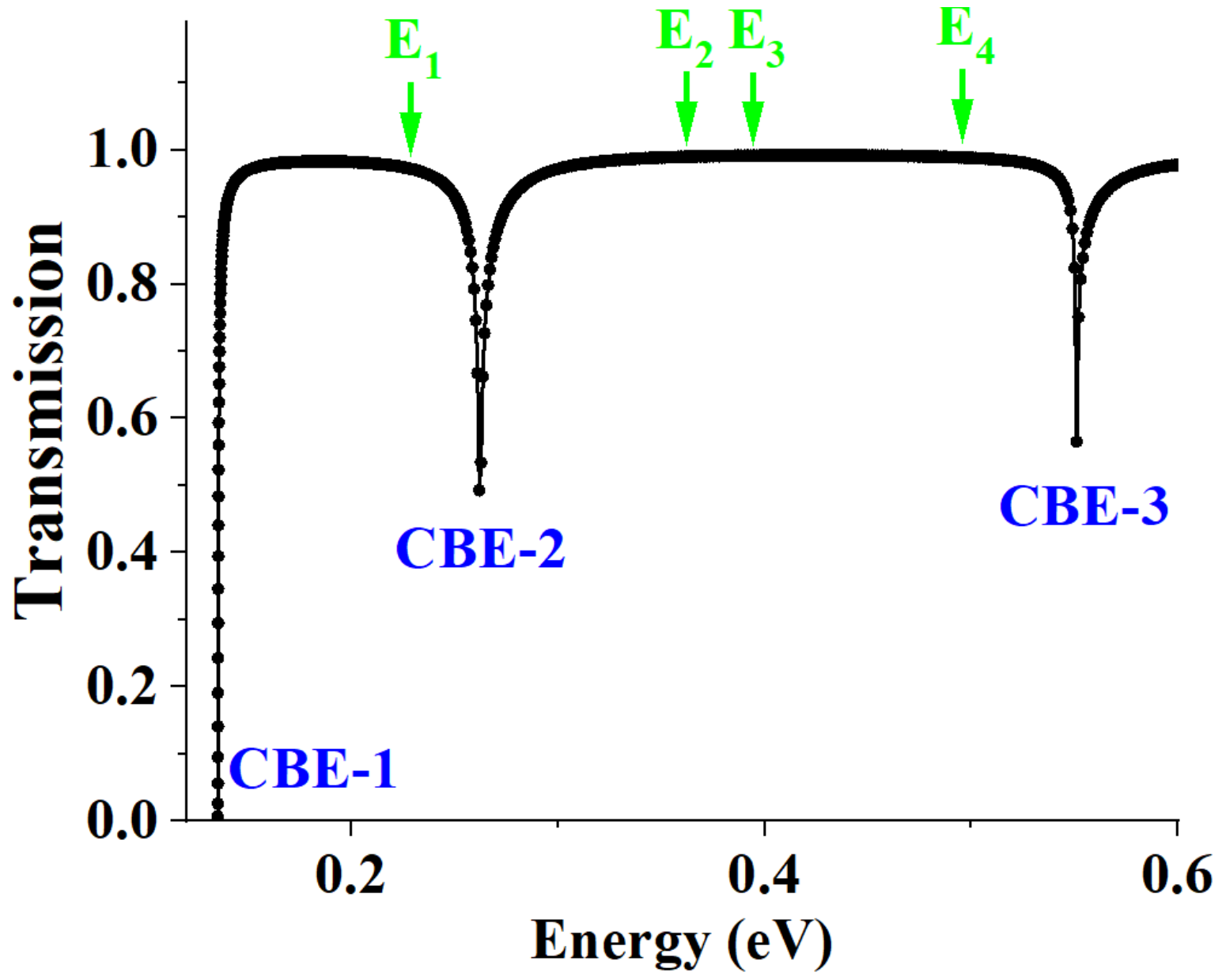

**Figure 5.** Transmission spectrum T vs. E for Structure X. T sets on at CBE-1 and rapidly approaches unity, with occasional dips occurring when E crosses a band edge; these transitions alter the coupling between incident and transmitted states, thereby modulating the transmission T.

To enable interface states to influence transmission, we introduce configurational modifications that hybridize these states with extended ones, transforming the localized states into transport-active, quasi-localized ones that couple to the continuum. Such quasi-localized states can manifest themselves as nontrivial variations, for example, possible Fano resonances, in the transmission spectrum.

The modified configuration, Structure Y, as shown in Figure 6, is designed for this purpose. First, we approximately duplicate the configuration of Structure X within Structure Y by increasing the transverse dimension to 47 hexagons. This structure comprises three parallel, horizontal stripes: top (15 hexagons), middle (17 hexagons), and bottom (15 hexagons). The middle stripe ($-y_c < y < y_c$) closely simulates Structure X; a linear-in-y potential, sgn($x$)V($y$), with an overall amplitude of 0.37 eV is applied across this stripe to approximately replicate the potential in Structure X. Second, a potential variation $\delta$V is introduced that lifts up the energy of the top and bottom stripes ($|y| > y_c$) by 0.5 eV relative to sgn($x$)V($y_c$) and sgn($x$)V($-y_c$), respectively. The resulting total potential V + $\delta$V is illustrated in Figure 6, shown for the drain side.

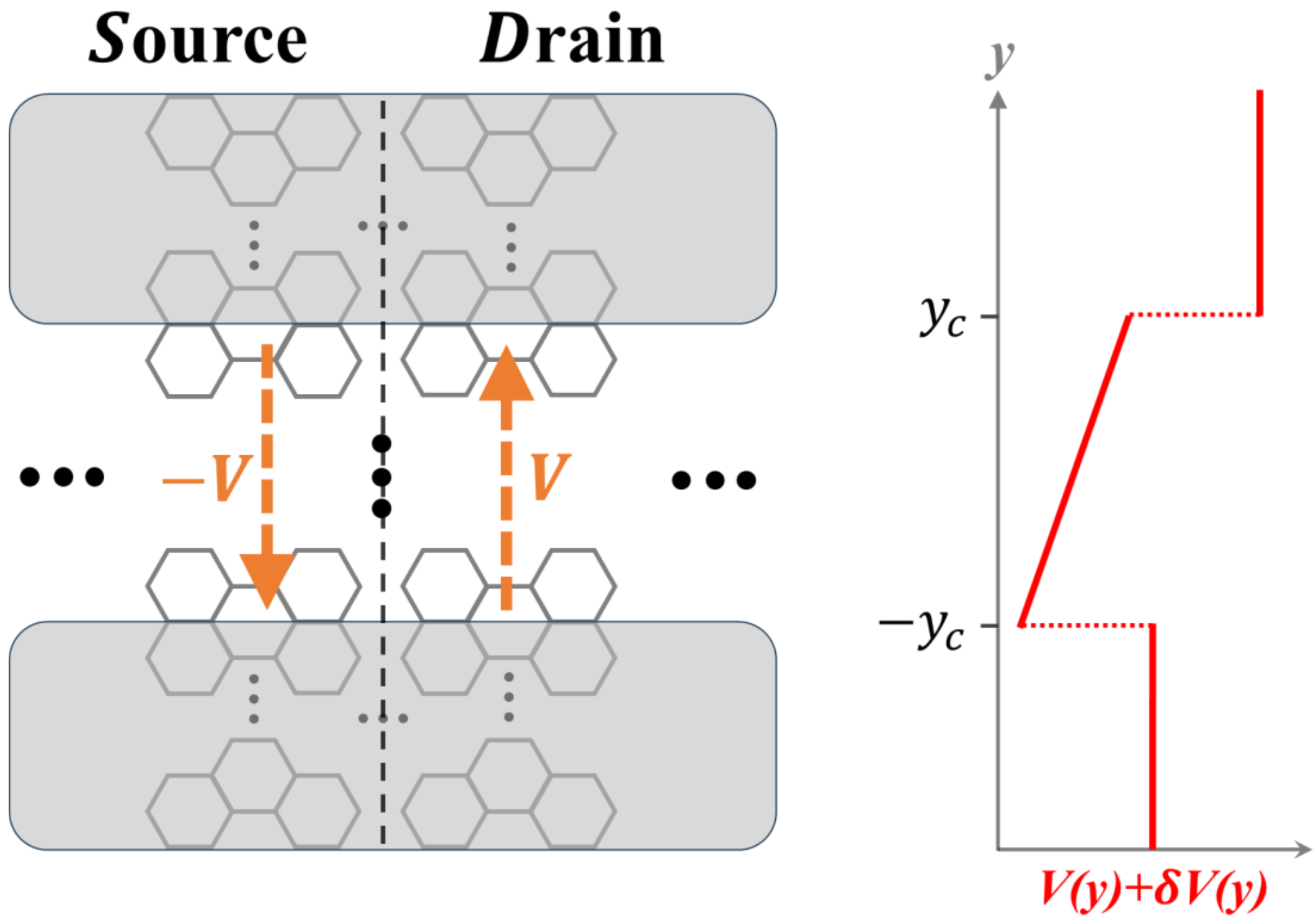

**Figure 6.** Schematics of Structure Y and the applied potential profile. The black dashed line indicates the interface at $x$ = 0. The structure consists of three parallel, horizontal stripes: top gray stripe ($y > y_c$), middle white stripe ($-y_c < y < y_c$), and bottom gray stripe ($y < -y_c$) with widths of 15, 17, and 15 hexagons, respectively. A transverse linear-in-y bias, sgn($x$)V($y$), with an overall amplitude of 0.37 eV is applied across the middle stripe. A nonlinear-in-y potential $\delta$V is introduced to lift up the energy of the top and bottom gray stripes ($y > y_c$ and $y < -y_c$) by 0.5 eV relative to sgn($x$)V($y_c$) and sgn($x$)V($-y_c$), respectively. The resulting overall potential, V + δV, is illustrated on the right for the drain side.

$\delta V$ induces a coupling between interface states and extended states. Consequently, it generates an interfacestate-assisted transmission channel in Structure Y, where an incoming electron hops from the incident state to an interface state, and then to the transmitted state. This occurs in addition to the primary "background channel", where the electron hops directly from the incident state to the transmitted state. The quantum interference between these two channels produces Fano resonances in the total transmission (T). As is well known, when the background transmission is near unity, a Fano resonance would appear as a "transmission zero", also known as “anti-resonance”—describing a nearly complete destructive interference—at the energy of the quasi-localized state [57–60].

Figure 7a presents T vs. E in Structure Y, showing the anti-resonance closest to the transmission on-set. The anti-resonance exhibits a line shape that approximately fits the Fano formula for complete destructive interference [57], i.e.,

$$T(E) \approx \frac{\epsilon^2}{1+\epsilon^2}, \epsilon = \frac{E - E_r}{\Gamma/2}, \tag{7}$$

where $E_r \approx 0.393$ eV is the quasi-localized state energy and $\Gamma \approx 6$ meV is the linewidth (1/lifetime). The presence of this anti-resonance strongly supports the existence of a quasi-localized state at $E_r$.

Further corroboration is provided in Figure 7b, which presents the un-normalized probability distribution obtained from the RGF calculation at E = $E_r$. The distribution exhibits a pronounced peak localized at the interface, accompanied by oscillatory leakage extending into the electrodes. Notably, the probability amplitude on the drain side decays asymptotically to zero, a behavior consistent with the vanishing transmission (T ≈ 0). While a detailed decomposition of the oscillatory features—which involve nanoribbon states with finite Re($k_x$)—is complicated, overall, the figure collectively demonstrates a clear signature of interface state-assisted transport: electrons hop from the incident state into the interface state and subsequently scatter into reflected or transmitted states.

From an experimental perspective, Figure 7 suggests that transport spectroscopy is a viable experimental method for directly probing valley-dependent topological interface physics in graphene nanoribbons. Specifically, Fano anti-resonances in transport measurements can serve to fingerprint interface states.

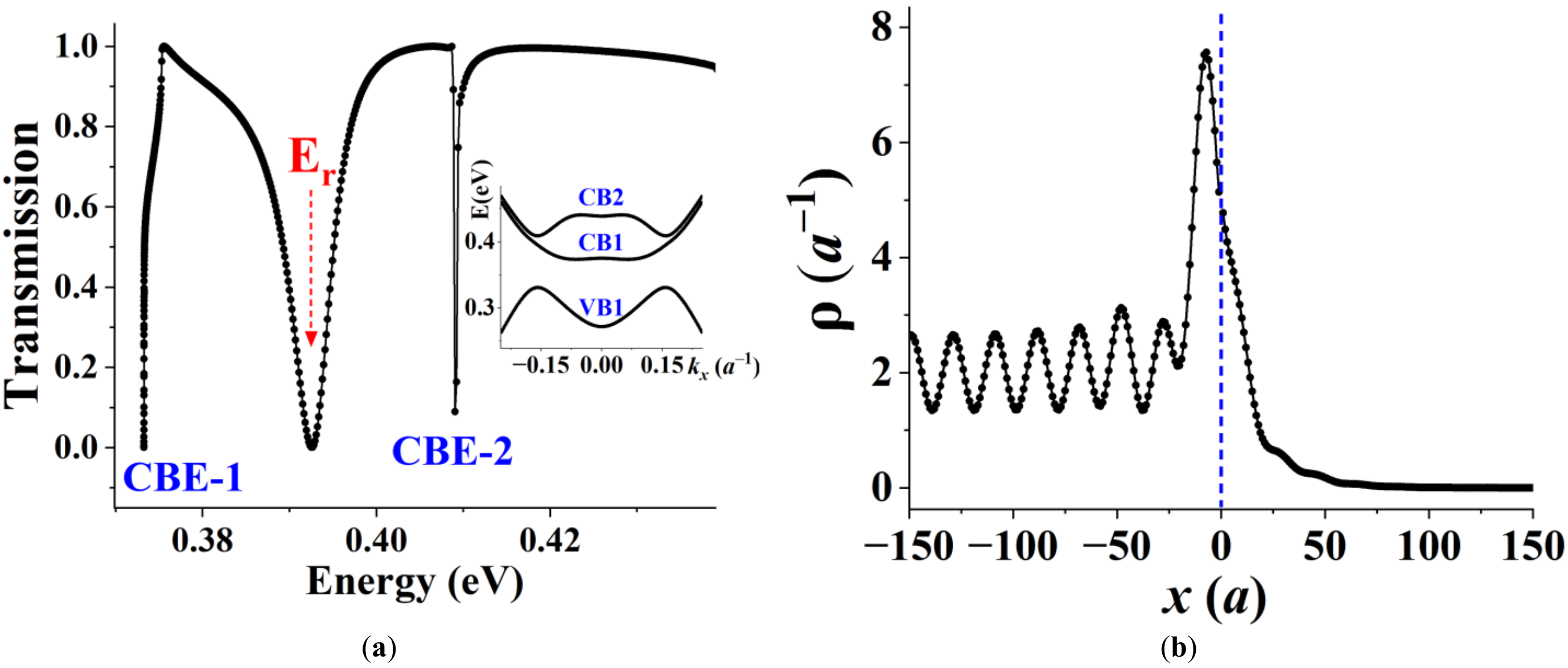

**Figure 7.** (**a**) Transmission spectrum of Structure Y, exhibiting a Fano transmission zero at the quasi-localized state energy $E_r \approx$ 0.393 eV. Inset: Band structure for the source and drain, showing one valence band (VB1) and two conduction bands (CB1, CB2), which aids in identifying band edge-induced discontinuities in the transmission. (**b**) Un-normalized 1D probability distribution calculated using the RGF method at E = $E_r$. The distribution exhibits a pronounced peak near the interface (dashed vertical line) with oscillatory leakage into the electrodes, confirming the character of a quasi-localized interface state.

We further examine the resilience of a quasi-localized interface state and its corresponding Fano signature. Figure 8 presents the transmission spectrum (T vs. E) for several cases, including the configuration studied in Figure 7a as a reference (black curve $Y^{l=0}_{V=0.37,\delta V=0.5}$). To simulate a more realistic structure, we insert an unbiased spacer region of thickness $l$, where $l$ = 10 nanoribbon unit cells, between the source and drain (blue curve $Y^{l=10}_{V=0.37,\delta V=0.5}$). In addition, we further investigate the effects caused by variations in the linear bias $V$, e.g., a reduction in the bias amplitude to 0.3 eV (red curve $Y^{l=0}_{V=0.3,\delta V=0.5}$), and by variations in the lift-up $\delta V$, e.g., a reduction to 0.4 eV (green curve $Y^{l=0}_{V=0.37,\delta V=0.4}$). In all these cases, the anti-resonance feature associated with an interface state remains clearly observable. The persistence indicates that the quasi-localized state and corresponding Fano signature are sufficiently robust, suggesting their experimental detectability under configurational fluctuations.

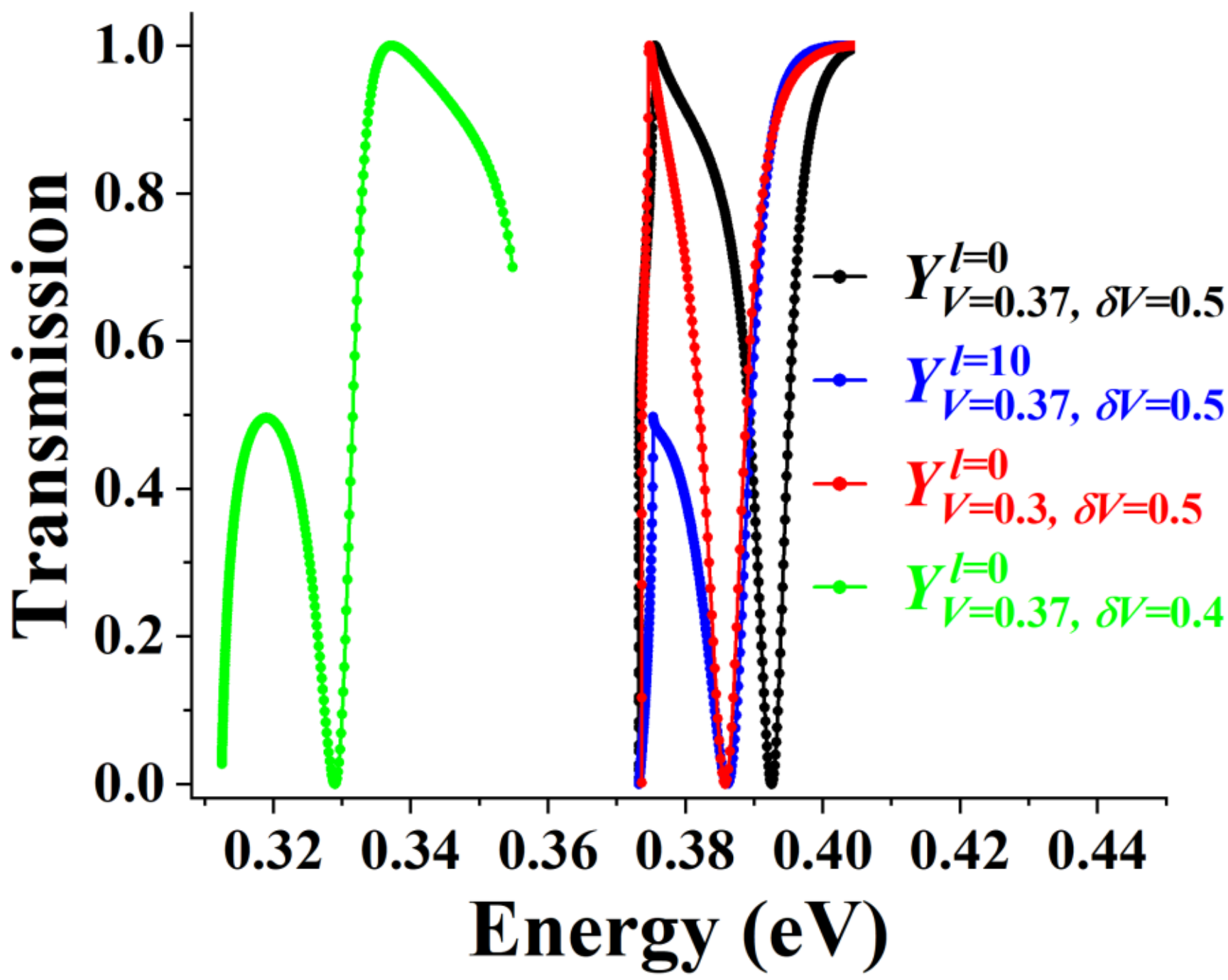

**Figure 8.** Transmission spectra under configurational variations. Black curve: Reference spectrum from Figure 7a. Blue curve: Spectrum with an unbiased spacer region of 10 nanoribbon cells inserted between the source and drain. Red curve: Spectrum with the linear bias (V) amplitude reduced to 0.3 eV. Green curve: Spectrum with the lift-up potential ($\delta$V) reduced to 0.4 eV. The Fano transmission zero persists in all cases, demonstrating robustness.

The above anti-resonant phenomenon may be utilized in device applications and open up new opportunities—for example, in valleytronic applications such as the selective filtering of carrier energy or switchable transport elements controlled electrically via V and $\delta$V. To illustrate such a potential, we investigate a combined structure below.

As depicted in Figure 9a, this combined device, designated as Structure Z, is formed of a symmetric resonant-tunneling structure (S1/B1/W1/B2), which transmits electrons within a narrow energy range, followed by Structure Y (S2/D2), which functions as an on/off switch for the transmission. In this study, the resonant-tunneling structure is taken to consist of an armchair GNR 13 hexagons wide, with two identical potential barriers (B1 and B2) both raised by 0.3 eV with respect to the incident electrode (S1) and the well (W1). Each of the B1, W1, and B2 regions is 20 nanoribbon unit cells long. It is well known that the electron transmission (T(E)) through such a symmetric resonant tunneling structure shows a series of unity peaks. Each peak signifies the existence of a quasi-bound electron state within the well (W1). The location of a peak reveals the energy of the corresponding state, $E_{quasi-bound}$, while the peak width reflects the rate of probability leakage of the quasi-bound electron through the barriers [61,62]. Structure Y here is taken to be spacer-free, with the source region (S2) 50 nanoribbon unit cells long. These parameters have been specifically tuned to ensure that the lowest $E_{quasi-bound}$ nearly matches the interface state energy $E_r$.

Figure 9b presents the transmission spectra for two combined structures. For comparison, the transmission for Structure Y alone, previously considered in Figure 8, is re-plotted here, as shown by the light-red curve ($Y^{l=0}_{V=0.3,\delta V=0.5}$). This curve exhibits a transmission zero near $E \approx 0.386$ eV for the structure under a bias amplitude of 0.3 eV.

The blue curve ($Z^{\delta V=0.5}_{V=0}$) depicts the transmission through the combined structure in the configuration where Structure Y is unbiased ($V = 0$). It shows a peak with an amplitude of ≈1, which primarily reflects the transport characteristics of the resonant-tunneling structure, with the peak location at $E \approx 0.388$ eV giving an estimate of the lowest $E_{quasi-bound}$ in W1.

The green curve ($Z_{V=0.3}^{\delta V=0.5}$) shows the transmission through the combined structure in the configuration, where S2 and D2 of Structure Y are oppositely biased with amplitude $V = 0.3$ eV. It shows that the biases nearly perfectly turn off the transmission.

Overall, Figure 9 demonstrates a proof-of-concept band-stop filter based on Structure Y, designed to block a specific range of electron energies. It also provides insight into the influence of Fabry–Pérot-type interference that may arise when Structure Y is connected in series with another quantum device. The results indicate that such interference between the two devices has only a marginal effect on the overall transport characteristics.

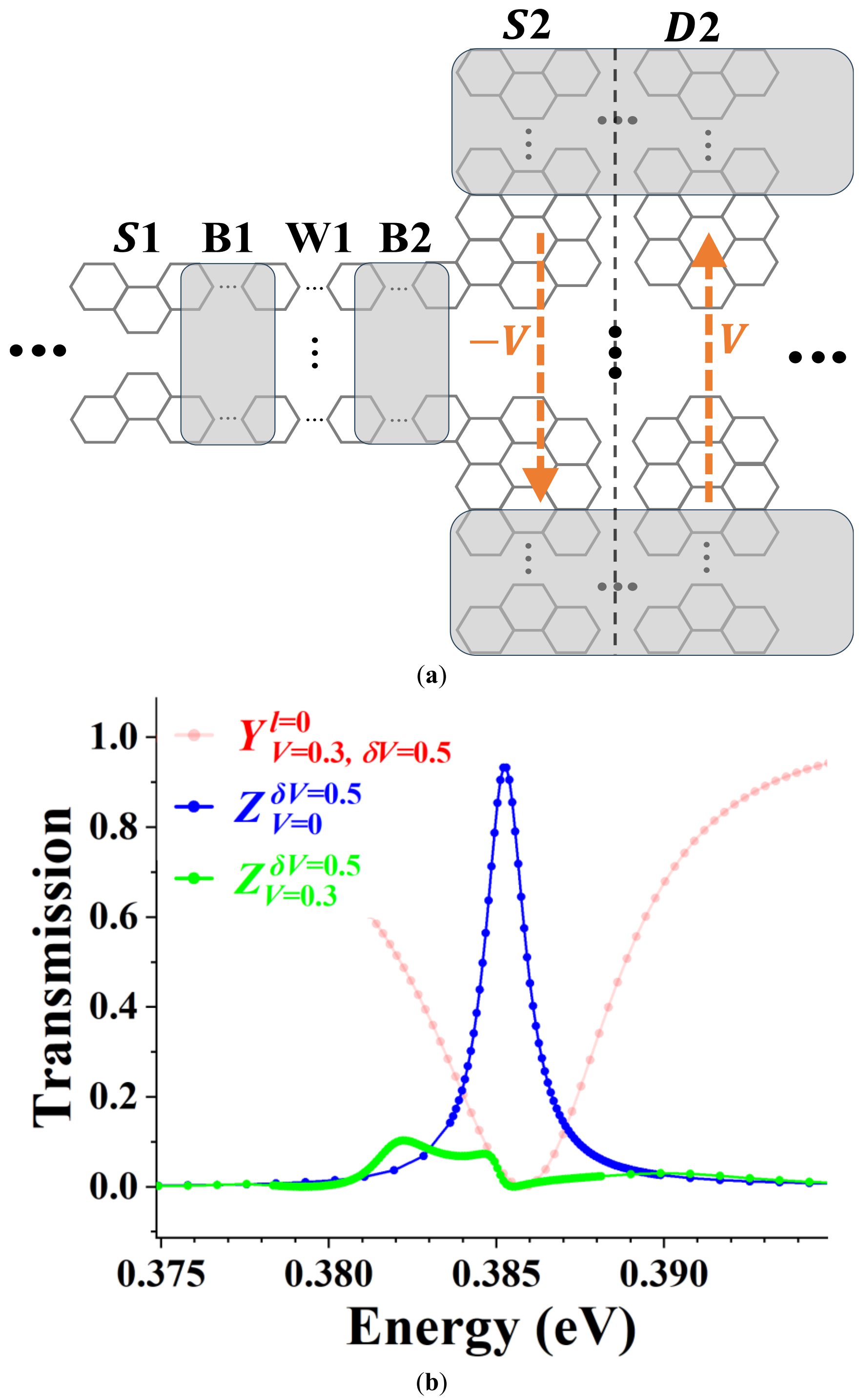

**Figure 9.** (**a**) Schematic of the combined structure, consisting of a resonant-tunneling structure (S1–B1–W1–B2) connected in series with Structure Y (S2–D2). W1 denotes the potential well, the two gray regions, B1 and B2, denote the potential barriers, and S1 denotes the incident electrode, in the resonant tunneling structure. V and -V indicate the in-plane, linear-in-y potentials applied to Structure Y, with the two gray stripes of Structure Y additionally subjected to the lift-up potential $\delta$V. (**b**) Transmission spectra. Light red curve: Transmission for Structure Y alone (from Figure 8). Blue curve: Transmission for the combined structure when Structure Y is unbiased (V = 0). Green curve: Transmission for the combined structure when Structure Y is biased (V = 0.3 eV).

## 5. Conclusions

Topological resilience offers a powerful means to mitigate configurational fluctuations in single-layer graphene-based devices. Motivated by this perspective, we investigated valley-dependent topological interface states in source/drain structures of armchair nanoribbons made of gapless single-layer graphene, featuring a bias-controlled valley polarization discontinuity at the interface.

We developed a theoretical framework to calculate nanoribbon complex band structures and construct interface state wave functions from exponentially decaying nanoribbon states, enforcing wave function continuity across the interface. This formulation was applied to a reference configuration, Structure X, where the bias-induced potential V is linear throughout the transverse dimension. Numerical results confirmed the existence of interface eigenstates with localized spatial probability distributions, residing both within the nanoribbon band gap and within the energy bands.

Regarding electron transport, we found that, in Structure X, the decoupling of interface eigenstates from extended ones prevents them from being transport-active. Consequently, we introduced a modified configuration, Structure Y, to enable hybridization between interface and extended states. The resulting quasi-localized states manifest as Fano anti-resonances in the transmission spectra. Notably, these features survive configurational fluctuations, including changes in the linear potential V and the lift-up potential $\delta$V, as well as the insertion of unbiased spacers.

We suggested transport spectroscopy as a feasible experimental method to directly probe valley-dependent topological interface physics in graphene nanoribbons, utilizing Fano anti-resonances to fingerprint interface states. Furthermore, we illustrated a proof-of-concept band-stop filter based on Structure Y. Our results show that, when connected in series with another quantum device, the transport characteristics of the combined structure are only marginally influenced by Fabry–Pérot-type interference between the two devices.

While anti-resonant phenomena have been investigated in mesoscopic non-topological systems—such as electronic waveguides with side cavities [59] and quantum wires side-coupled to quantum dots [60]—the unique topological resilience demonstrated here makes the present system particularly promising for realizing similar functionalities. More broadly, for nanoscale devices in gapless single-layer graphene, the demonstrated robustness against configurational fluctuations supports the exploitation of valley-dependent topological physics as a viable pathway toward practical implementation.

**Funding:** This research was funded by National Science and technology council (Taiwan), grant number NSTC-110-2112- M-007-038.